\documentclass[11pt,article,aps,prd,groupedaddress,preprint,eqsecnum, nofootinbib]{revtex4}
\usepackage{graphicx,epsf,amssymb,amsbsy,amsfonts,amssymb,amsmath,setspace}

\def\be{\begin{equation}}
\def\ee{\end{equation}}

\def\bg{\bar{g}}
\def\beq{\begin{eqnarray}}\def\eeq{\end{eqnarray}}
\def\ba#1\ea{\begin{align}#1\end{align}}
\def\bg#1\eg{\begin{gather}#1\end{gather}}
\def\bm#1\em{\begin{multline}#1\end{multline}}
\def\bmd#1\emd{\begin{multlined}#1\end{multlined}}

\def\({\left(}
\def\){\right)}
\def\[{\left[}
\def\]{\right]}

\begin{document}
\hfuzz 12pt

\title{Black Hole Singularity, Generalized (Holographic) $c$-Theorem and Entanglement Negativity}
\author{Shamik Banerjee}
\affiliation{Kavli Institute for the Physics and Mathematics of the Universe, The University of Tokyo,\\
5-1-5 Kashiwa-no-Ha, Kashiwa City, Chiba 277-8568, Japan }

\email{banerjeeshamik.phy@gmail.com}

\author{Partha Paul}

\affiliation{Institute of Physics, \\ Sachivalaya Marg, Bhubaneshwar, India-751005}

\email{pl.partha13@iopb.res.in}  

\begin{abstract}
In this paper we revisit the question that in what sense empty $AdS_{5}$ black brane geometry can be thought of as RG-flow. We do this by first constructing a holographic $c$-function using causal horizon in the black brane geometry. The UV value of the $c$-function is $a_{UV}$ and then it decreases monotonically to zero at the curvature singularity. Intuitively, the behavior of the $c$-function can be understood if we recognize that the dual CFT is in a thermal state and thermal states are effectively massive with a gap set by the temperature. In field theory, logarithmic entanglement negativity is an entanglement measure for mixed states. For example, in two dimensional CFTs at finite temperature the renormalized entanglement negativity of an interval has UV (Low- T) value $c_{UV}$ and IR (High-T) value zero. So this is a potential candidate for our $c$-function. In four dimensions we expect the same thing to hold on physical grounds. Now since the causal horizon goes behind the black brane horizon the holographic $c$-function is sensitive to the physics of the interior. Correspondingly the field theory $c$-function should also contain information about the interior. So our results suggest that high temperature (IR) expansion of the negativity (or any candidate $c$-function) may be a way to probe part of the physics near the singularity. Negativity at finite temperature depends on the full operator content of the theory and so perhaps this can be be done in specific cases only.

The existence of this $c$-function in the bulk is an extreme example of the paradigm that space-time is built out of entanglement. In particular the fact that the $c$-function reaches zero at the curvature singularity correlates the two facts : loss of quantum entanglement in the IR field theory and the end of geometry in the bulk which in this case is the formation of curvature singularity.



%
\end{abstract}

\preprint{}
\maketitle
\tableofcontents

\section{Introduction}

In AdS-CFT black brane is a thermal state of the boundary conformal field theory living on the Minkowski space-time. This is not a relevant deformation of the CFT Hamiltonian and there is no renormalization group flow in the ordinary sense. Therefore the question of the existence of a $c$-function, in the sense of Zamolodchikov \cite{Zam, Cappelli:1990yc,Cardy:1988cwa}, does not naturally arise in this situation. Moreover Zamolodchikov $c$-function is constant at a fixed point and independent of the state of the CFT. The purpose of this note is to point out that AdS-CFT duality and the thermodynamic nature of classical gravity allows us to introduce a generalized notion of $c$-function, at least for large-$N$ theories with classical gravity dual. This generalized $c$-function cannot be interpreted as an off-shell central charge. Rather it can be interpreted as a measure of quantum entanglement that exists at different energy scales in the given state. We will construct this $c$-function holographically when the CFT is in thermal state and the gravity dual is an empty black brane geometry. We focus on four dimensional field theories only. Our choice of the thermal state is motivated by the fact that the gravity dual has a curvature singularity and the Lorentz invariance is broken everywhere except near the UV boundary of AdS. So it can teach us some lessons about RG-flow interpretation of more general geometries.




\textbf{Throughout this paper we will assume that the bulk theory is Einstein gravity coupled minimally to a set of matter fields.}

\section{Holographic View}


The holographic picture is based on the fact that the gravity dual of $c$-theorem is the second law of causal horizon thermodynamics in asymptotically AdS spaces \cite{sb, Banerjee:2015uaa}. In a nutshell, second law for causal horizons say that if we consider the future bulk light-cone of a boundary point then the expansion of the null geodesic generators of the light-cone is negative \cite{Gibbons:1977mu, Jacobson:2003wv}. Now one can assign Bekenstein-Hawking entropy to the causal horizon. The fact that the expansion is negative then implies that as we move away from the boundary the entropy density decreases monotonically. This is essentially holographic $c$-theorem \cite{Akhmedov:1998vf,Freedman:1999gp,cthor, deBoer:1999xf, ctheorems, Myers:2010xs} if we specialize to a domain-wall geometry. The bulk future light-cone interpolates between the UV-AdS and the IR-AdS and the monotonically decreasing Bekenstein-Hawking entropy density gives the holographic $c$-function \cite{sb, Banerjee:2015uaa}.

If we focus on domain-wall geometry then the second law has the interpretation of holographic $c$-theorem. But what about other asymptotically AdS (AAdS) geometries ? Second law of causal horizon thermodynamics holds in any AAdS geometry and in fact holographic RG \cite{deBoer:1999xf} applies to any such setup. It has been argued that the holographic RG in the bulk is dual to the Wilsonian RG in the boundary \cite{Heemskerk:2010hk,Faulkner:2010jy,Elander:2011vh, Radicevic:2011py}. So is it possible to associate a notion of irreversibility to any classical AAdS geometry ? It seems that the existence of second law for classical gravity allows us to do precisely this thing. In the field theory side its interpretation will require us to generalize the concept of Zamolodchikov-type $c$-function. We will have almost nothing to say on it in this paper. 

  To gain some experience with such generalized $c$-functions we will work out a reasonably simple but interesting example of empty black brane geometry in AdS$_{5}$ \footnote{Construction of holographic $c$-function by viewing the black brane background as RG flow, was also considered in \cite{Paulos:2011zu,Cremonini:2013ipa,Liu:2012wf}. $c$-function for attractor flows were considered in \cite{Goldstein:2005rr}. }.  It has curvature singularity hidden behind the black brane horizon and the geometry is not Lorentz invariant except near the AdS$_{5}$ boundary. The $c$-function we construct is just the Bekenstein-Hawking entropy density of a causal horizon in the black brane geometry \cite{sb, Banerjee:2015uaa}. The causal horizon originates at some point of the AdS boundary and terminates at the curvature singularity. Nothing depends on the choice of the boundary point where the causal horizon originates because of space-time translation invariance. Second law guarantees that our function monotonically decreases as we move away from the boundary along the null geodesic generators of the causal horizon. We will see that the $c$-function monotonically decreases from $a_{UV}$ to zero at the curvature singularity.

\subsection{Calculation and Results}














The causal horizon is just the future bulk light-cone of a boundary point. We take the boundary point, $p$, to have coordinates, $x^\mu = z =0$.

The metric of the five-dimenional black brane is, 

\begin{equation}\label{bb}
ds^2 =  \frac{1}{z^2} \Big[ -(1- z^4) dt^2 + \frac{dz^2}{1-z^4} + d\vec x^2 \Big]
\end{equation}

Our job is to construct the ingoing null geodesics in this geometry which originate from the boundary point $p$. 



\begin{figure}[htbp]
\begin{center}
  \includegraphics[width=15cm]{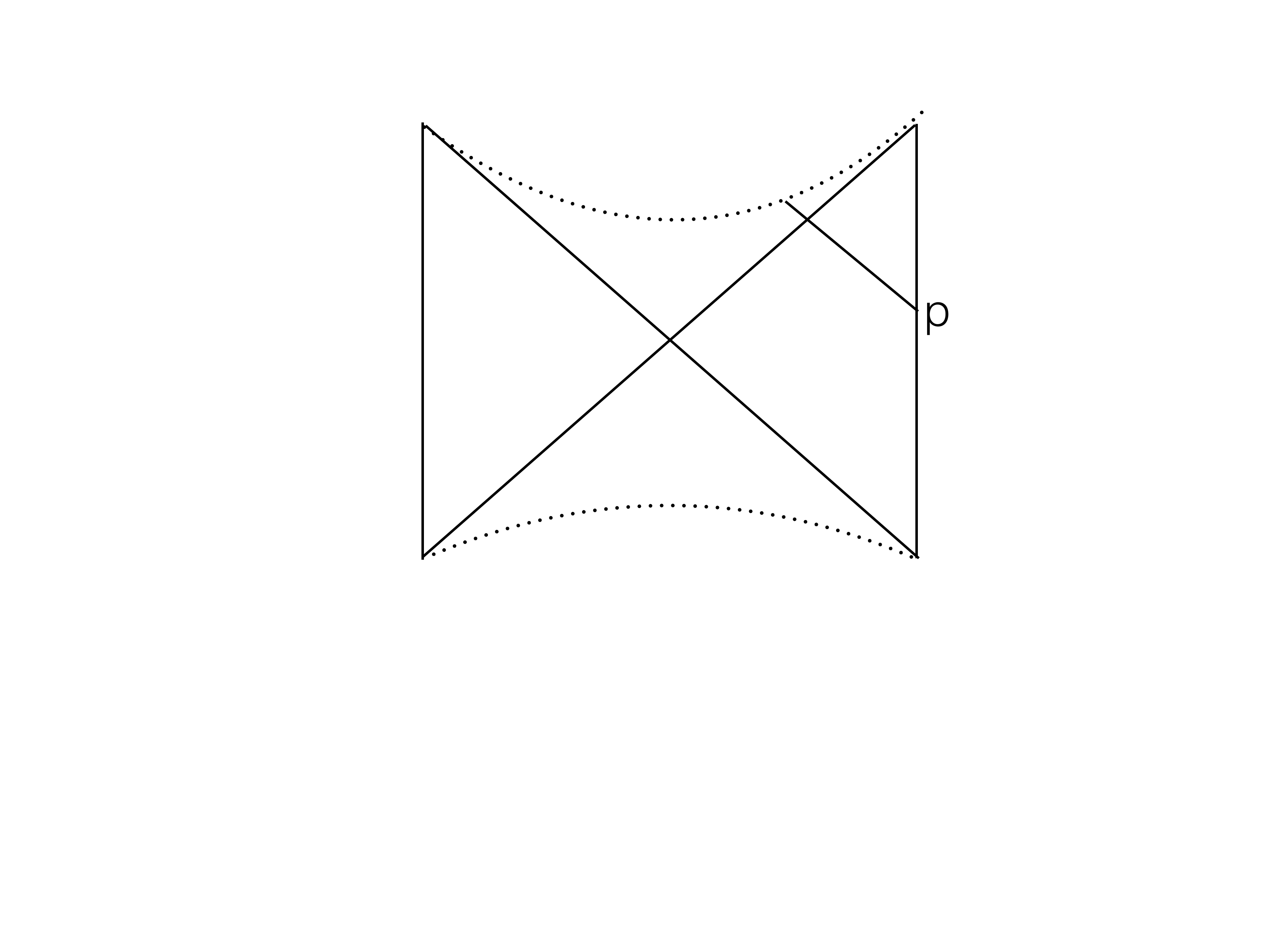}
\end{center}
\vspace{-11em}
\caption{Penrose diagram of the maximally extended AdS$_5$ black brane \cite{Fidkowski:2003nf}. We have shown only the radial null geodesic coming out from the boundary point $p$. Other non-radial null geodesics from $p$ are not shown here. In this paper we do not need the setup of the two sided black brane. We have drawn it for the sake of completeness.}
\end{figure}

Let us define the ingoing Eddington-Finkelstein coordinate as, $ v = t + z^{*} $ where $ z^{*} = \frac{1}{2} \tan^{-1}z + \frac{1}{4} \log \left( \frac{1+z}{1-z} \right) $, so that the metric takes the form

\begin{equation}
ds^2 =   \frac{1}{z^2}\Big[ -(1- z^4) dv^2 -2dvdz + d\vec{x}^2 \Big] 
\end{equation}

There is no singularity at the horizon, $z=1$, and so we can follow the null geodesics all the way to the curvature singularity at, $z=\infty$. Since we want to find out the null geodesics we can as well work with the conformally transformed metric given by, 

\begin{equation}
d\tilde s^2 = -(1- z^4) dv^2 -2dvdz + d\vec{x}^2 
\end{equation}

Let $\lambda$ denote the affine parameter along a null geodesic in the conformally transformed metric $d\tilde s^2$. \footnote{We could not solve the equation for $z(\lambda)$ by using the affine parameter corresponding to the original geometry. But this does not affect the physics. This is just a change in scheme. It will of course be better to solve this in terms of the original affine parameter }


 So we have,

\begin{equation}\label{NGC}
\tilde g_{AB} \frac{dx^A}{d\lambda} \frac{dx^B}{d\lambda} = 0
\end{equation}

We also have four conserved charges corresponding to the translations in $v$ and the $x^{i}$'s. 

\begin{align}\label{CC}
\begin{split}
-(1-z^4) \frac{dv}{d\lambda} &= - E + \frac{dz}{d\lambda} , \\
 \frac{dx^i}{d\lambda} &= -p^i ,  
 \end{split}
\end{align}

where $i=1,2,3$. $E$ and $\vec p$ are the conserved charges along a null geodesic. Here we are assuming that the affine parameter $\lambda$ increases as we move away from the boundary at $z=0$. We will be working with the future bulk light-cone of the boundary point $p$ and so with our convention for the affine parameter, $\frac{dt}{d\lambda}\ge 0$. So $E\ge 0$. 

Now using ($\ref{NGC}$) and ($\ref{CC}$) we get,

\begin{equation}
\Big(\frac{dz}{d\lambda}\Big)^2 = E^2 - {p}^{2} (1-z^4)
\end{equation}

So we can see that the null geodesics which can reach the boundary point must satisfy the constraint,  $E^2 - p^2\ge 0$. This constraint together with the constraint $E \ge 0$, allow us to parametrize the conserved charges as,

\begin{align}
\begin{split}
E &= \alpha \cosh\eta \\
p^i &= \alpha \sinh\eta \ \hat n^i \\
\end{split}
\end{align}

where $\alpha> 0$, $0\le\eta\le\infty$ and $\hat n$ is a unit vector in $R^3$. Now it is easy to see that $\alpha$ is redundant because it can be absorbed by an affine reparametrization, $\lambda\rightarrow\alpha\lambda$. Therefore we will set $\alpha=1$.


So the equation for $z$ simplifies to,

\begin{equation}
\frac{dz}{d\lambda} = \sqrt{1+ z^4 \sinh^4\eta}
\end{equation}

We have chosen the positive root because our convention is $\frac{dz}{d\lambda}\ge 0$. So we can write,

\begin{equation}\label{sing}
\lambda = \int_{0}^{z} \frac{dz'}{\sqrt{ 1+ z'^4 \sinh^2\eta}}
\end{equation}

where the boundary condition, $z(0)=0$ has been imposed. 


















The solution of this equation is, \footnote{We are using the convention of Gradshteyn and Ryzhik. In Mathematica, $\frac{1}{\sqrt 2}$ in the argument of $cn$ should be replaced by $\frac{1}{2}$.}

\begin{equation}
z^2(\lambda) = \frac{1}{\sinh\eta} \frac{1-cn(2\lambda\sqrt{\sinh\eta},1/\sqrt{2})}{1+cn(2\lambda\sqrt{\sinh\eta},1/\sqrt{2})}
\end{equation}

where $cn$ is one of the Jacobian elliptic functions. Its properties are well studied although a closed form expression in terms of elementary functions does not exist. 

Given the solution for $z(\lambda)$ we can in principle determine $v(\lambda)$ from (\ref{CC}), but we were unable to do so in any convenient way. In any case the complete set of solutions can be written as,

\begin{align}\label{nh}
\begin{split}
z(\lambda,\eta) &=\sqrt{ \frac{1}{\sinh\eta} \frac{1-cn(2\lambda\sqrt{\sinh\eta},1/\sqrt{2})}{1+cn(2\lambda\sqrt{\sinh\eta},1/\sqrt{2})}} \\
v(\lambda,\eta) &= \int_{0}^{\lambda} d\lambda' F(\lambda',\eta) \\
x^i(\lambda,\eta,\hat n^i) &= -\lambda  \sinh\eta \ \hat n^i
\end{split}
\end{align}

where we have defined, 
\begin{equation}
F(\lambda,\eta) = \sinh^2\eta \frac{(1+ cn)^2\cosh\eta - \sqrt 2 \sqrt{1+cn^2} (1+ cn)}{(1+cn)^2 \sinh^2\eta -(1-cn)^2} 
\end{equation}

and $cn\equiv cn(2\lambda \sqrt{\sinh\eta},\frac{1}{\sqrt 2})$. We have imposed boundary conditions such that, $z(0,\eta) = v(0,\eta)= x^i(0,\eta,\hat n^i) =0$ for all values of $\eta$ and $\hat n^i$. This corresponds to the fact that the null geodesics are all coming out of the point $p$ with coordinates $x^i= v=z=0$. Note that $v=t$ at the boundary $z=0$.

For any fixed values of $\eta$ and $n^i$, the above equation ($\ref{nh}$) reduces to the equation of the null geodesic parametrised by the affine parameter $\lambda$ and coming out of the fixed boundary point $p(x^i=t=z=0)$. As we vary $\eta$ and $n^i$, we scan over all the geodesics coming out of the point $p$. All these null geodesics form a null hyper surface whose parametric equation is given by ($\ref{nh}$). The intrinsic coordinates on the null hyper surface are ($\lambda,\eta,\hat n^i$). ($\eta,\hat n^i$) are comoving coordinates along a null geodesic parametrised by $\lambda$. This null hyper surface is the sought for bulk future light-cone or the past causal horizon of the point $p$.

Our next job is to find out the induced metric on the null-hypersurface ($\ref{nh}$). To find out the induced metric we have to use the original black brane metric ($\ref{bb}$). Using this we get,

\be\label{im}
ds_{ind}^2 = \frac{1}{z^2} \Big[-(1-z^4)\Big(\frac{\partial v}{\partial\eta}\Big)^2 - 2 \frac{\partial v}{\partial\eta}\frac{\partial z}{\partial \eta} + \lambda^2 \cosh^2\eta \Big] \\ d\eta^2 + \frac{1}{z^2} \lambda^2 \sinh^2\eta d\Omega_2^2
\ee

where $d\Omega_2^2$ is the metric of a unit two-sphere parametrised by $\hat n^i$. The induced metric is degenerate as it should be because ($\ref{nh}$) is a null-hypersurface.  ($\ref{im}$) is the metric on a $\lambda=constant$ space-like slice of the causal horizon ($\ref{nh}$), parametrised by the coordinates ($\eta,\hat n^i$). 

The volume form can be written as,

\be
dV_{ind} = c(\lambda,\eta) \ dV_{H^3}
\ee

where we have defined,

\be
c(\lambda,\eta)=\frac{\lambda^2}{z^3} \sqrt{\Big[-(1-z^4) \Big(\frac{\partial v}{\partial \eta}\Big)^2 - 2 \frac{\partial v}{\partial \eta} \frac{\partial z}{\partial \eta} + \lambda^2 \cosh^2\eta \Big]}
\ee

$dV_{H^3}$ is the volume form on a unit three dimensional hyperbolic space given by, 

\begin{align}
\begin{split}
ds_{H^3}^2 &= d\eta^2 + \sinh^2\eta d\Omega_2^2 \\
dV_{H^3} &= \sinh^2\eta \sin\theta d\eta d\theta d\phi
\end{split}
\end{align}

where we have parametrised $\hat n^i$ as ($\sin\theta\cos\phi, \sin\theta\sin\phi, \cos\theta$). The fact that $c$ is a function only of $\lambda$ and $\eta$ is a consequence of the rotational symmetry of the metric. In the more standard domain-wall geometry $c$ is function only of $\lambda$ because of the Lorentz invariance of the metric. In the black brane geometry Lorentz invariance is broken down to the spatial rotation group and so the $\eta$ dependence is non-trivial. 

Now second law for causal horizons is the statement that,

\be\label{sl}
\boxed{\frac{\partial}{\partial\lambda} \Big |_{\eta} c(\lambda,\eta)  \le 0}
\ee

Here we have used the fact that $dV_{H^3}$ is a comoving volume element and $\eta$ is a comoving coordinate i.e, $\eta$ is constant along a null geodesic generator of the causal horizon.

The Bekenstein-Hawking entropy density associated to the volume element $dV_{ind}$ is,

\be
dS_{BH} = \frac{dV_{ind}}{4G_N} = \frac{c(\lambda,\eta)}{4G_N} dV_{H^3}
\ee

We can put in the AdS radius $L$ by replacing $dV_{ind}\rightarrow L^3 dV_{ind}$. This gives,

\be
dS_{BH} = \frac{dV_{ind}}{4G_N} = \frac{L^3}{4G_N} c(\lambda,\eta) dV_{H^3}
\ee

So our $c$-function is,

\be
\boxed{c_{\eta}(\lambda) = \frac{L^3}{4G_N} c(\lambda,\eta)}
\ee


We get a family of $c$-functions parametrised by $\eta$ (Fig-2). We check in the appendix using perturbation theory for small $\lambda$ that $c(\lambda,\eta)\rightarrow 1$ as $\lambda\rightarrow 0$ for \textbf{all} values of $\eta$ i.e, $c(0,\eta)=1$. It will be true for any AAdS geometry, not just the black brane. Note that $\lambda=0$ is the  AdS boundary and $\lambda$ increases as we move away from the boundary along the null geodesics. So for any fixed value of $\eta$ the $c$-function $c_{\eta}(\lambda)$ starts at the UV value $a_{UV}$ and decreases monotonically as a result of the second law ($\ref{sl}$). It turns out that in the case of the black brane the $c$-function becomes zero at the curvature singularity for \textbf{all} values of $\eta$. So for black brane in five dimensions, the $c$-function monotonically decreases from the UV central charge to zero at the curvature singularity. It does not show any characteristic behavior while crossing the black brane horizon.

\begin{figure}[htbp]
\begin{center}
  \includegraphics[width=10cm]{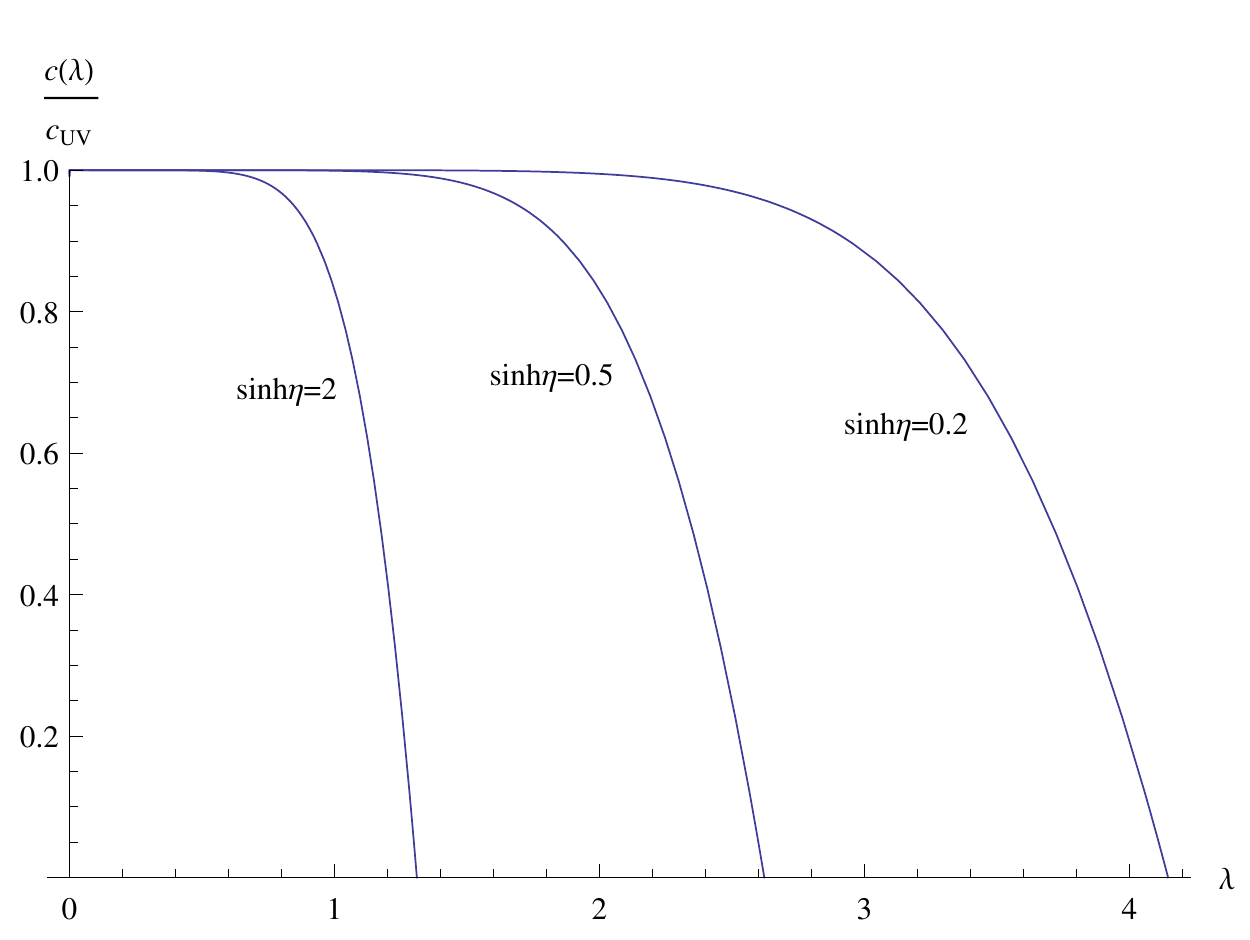}
\end{center}
\caption{We have plotted the $c$-function for three different values of $\eta$. All of them start at the UV value $a_{UV}(=c_{UV})$ and monotonically decreases to zero at the curvature singularity. The values of $\lambda$ at the singularity for different values of $\eta$ can be obtained from ($\ref{sing}$) by setting $z=\infty$. Of course from a physical point of view going to the singularity with GR is meaningless. But if we forget about any stringy physics for the time being, then as a classical theory GR holds everywhere except at the singularity.  }
\vspace{0em}
\end{figure}

We would like to emphasise that the fact that we have obtained a family of $c$-functions parametrized by $\eta$, instead of just one, is no cause for concern. $c$-function is not unique. For example in two dimensions one can construct the standard Zamolodchikov $c$-function \cite{Zam} and also the entanglement entropy $c$-function due to Casini and Huerta \cite{Cardy:1988cwa}. It is know that they are not the same, but they both monotonically interpolate between the UV and the IR central charges. In fact if we can construct one $c$-function then we can construct an infinite family all of which contain the same physical information \cite{Cappelli:1990yc}. 

The plot of the $c$-function in Fig-2 shows that it is not stationary at the singularity. This is not a problem because strictly speaking the function is not analytic there. We do not know how to extend the function beyond the singularity. But the fact that it is zero at the singularity shows that the flow comes to an end at the singularity. The $c$-function is an element of area of the causal horizon and so it is positive semidefinite by construction. So the flow saturates the lower bound at the singularity. This is similar to what sometimes happens in case of the $c$-function constructed out of entanglement entropy. For example in three dimensions, the entropic $c$-function for a massive scalar is not stationary at the UV fixed point \cite{Klebanov:2012va}. This is attributed to the fact that a scalar field with negative mass squared is pathological and the entropic $c$-function knows about that \cite{Casini:2015woa}.  Also another is that in our case the geometry is not Lorentz-invariant anywhere except near the boundary and so our standard intuition about $c$-function may need some modification. 

Before we conclude we would like to mention an important point. In Einstein gravity one cannot really distinguish between the $a$ and $c$ central charges. In order to do that one has to include higher-derivative terms in the bulk gravity action. In the presence of higher derivative terms instead of Bekenstein-Hawking entropy we have to use the entropy expression which satisfies the second law in the bulk  and reduces to the Wald entropy when evaluated on a Killing horizon \cite{Wald}. If we do this we will recover the $a$-charge at the asymptotic UV boundary as was shown in \cite{Banerjee:2015uaa}. That means the $a$-function will start decreasing from $a_{UV}$. The important point is the fate of this $a$-charge in the deep IR i.e, when the causal horizon reaches the singularity. We expect it to go to zero because the thermal state has a finite correlation length even in the presence of the higher-derivative terms, but proving this in general seems to be a complicated thing.


\section{Towards A Physical Interpretation}

Empty black brane in AdS is dual to a thermal state of the boundary conformal field theory (CFT) \cite{Witten:1998zw}. This is not a relevant deformation of the CFT Hamiltonian and there is no renormalization group (RG) flow in the ordinary sense. So it is unlikely that the holographic $c$-function is an off-shell central charge.  To make further progress, it will be useful to take note of the fact that a thermal state is effectively massive with a gap set by the temperature. There is a finite correlation length of the order of inverse temperature. The IR behavior of the holographic $c$-function that we have constructed shows the presence of this effective mass gap. It is monotonically decreasing from the central charge of the UV-CFT, $a_{UV}$, to zero at the curvature singularity which is in the deep IR and space-time ends there. \textit{Therefore the causal-horizon $c$-function faithfully quantifies the amount of pure quantum correlation or the effective number of "quantum degrees of freedom" that exists at different scales in the thermal state.} 


Can this be related to renormalized entanglement entropy in the boundary theory ? First of all space-like slices of the causal horizon are not in general extremal surfaces in the bulk \cite{Ryu:2006bv}. In the field theory side, suppose we consider a ball in $R^3$ of radius $R$. This is our subsystem for which we want to compute the renormalized entanglement entropy \cite{Liu} when the field theory is in the thermal state. Since the the theory is scale invariant the renormalized entanglement entropy will have the functional form $S_{REE}(R T)$, where $T$ is the temperature. It is known that as $T\rightarrow 0$, $S_{REE}\rightarrow a_{UV}$ \cite{Liu}. This matches with the behavior of our $c$-function in the same limit. In the opposite limit of $T\rightarrow\infty$ on the other hand the renormalized entanglement entropy $S_{REE}$ is nonzero and dominated by thermal entropy of the system \cite{Liu}. This does not match with the behavior of the $c$-function. This is not surprising because entanglement entropy is not an entanglement measure in a mixed state. In the high temperature limit it is contaminated by classical correlations and fails to capture the quantum part, which should go to zero. On the contrary the behavior of the causal horizon $c$-function shows that it is sensitive only to quantum correlations. Is there a candidate for such a quantity in the field theory ? 

\subsection{Is Finite Temperature Entanglement Negativity A Generalized $c$-function ?}

As we have discussed entanglement entropy at finite temperature is not a candidate for this generalized $c$-function because it is not an entanglement measure in a mixed state. One such measure which can be calculated in field theory is entanglement negativity \cite{Vidal:2002zz,Vidal:1998re,Calabrese:2012ew,Calabrese:2014yza, zol, Rangamani:2014ywa,Perlmutter:2015vma,Kulaxizi:2014nma}. Entanglement negativity was studied from a holographic point of view in \cite{Rangamani:2014ywa}, but to the best of our knowledge a geometric prescription of computing this in gravity does not exist so far. 

Entanglement negativity at finite temperature in a two dimensional CFT was computed in \cite{Calabrese:2014yza} \footnote{See also \cite{zol}.} . They calculated this for a single interval of length $L$ when the total system lives on an infinite line and the temperature is $T=\beta^{-1}$. 
In this case the answer is given by,

\be\label{N}
E = \frac{c}{2} \ln \Big[\frac{\beta}{\pi a} \sinh\Big( \frac{\pi L}{\beta}\Big) \Big] - \frac{\pi c L}{2\beta} + f(e^{-\frac{2\pi L}{\beta}}) + 2\ln c_{\frac{1}{2}}
\ee

where $a$ is the short distance cutoff, $c$ is the central charge of the CFT and $c_{\frac{1}{2}}$ is a constant. $f(x)$ is a universal scaling function which depends on the full operator content of the CFT  such that $f(1)=0$ and $f(0)=$constant. Given this we can calculate its value in the UV and the IR. UV is the region where $\beta>> L$ and we get,

\be
E_{UV} = \frac{c}{2} \ln\frac{L}{a} + 2\ln c_{\frac{1}{2}}
\ee

which is the correct zero temperature result. Similarly in the IR, $a<<\beta<< L$ and we get,

\be
E_{IR} = \frac{c}{2} \ln\frac{\beta}{2\pi a} + f(0) + 2\ln c_{\frac{1}{2}}
\ee

So in the IR this becomes a non-universal constant independent of the length $L$ of the subsystem \cite{Calabrese:2014yza}. The second term in ($\ref{N}$) is very important in the high temperature limit because it cancels the contribution to the negativity which is extensive in $L$. This is the principal difference from entanglement entropy which is useful for us. Now if we define a renormalized negativity, $E_{R}$, just like renormalized entanglement entropy \cite{Casini:2004bw,Liu}, as, 

\be
E_{R} = L \frac{d}{dL} \Big|_{\beta} E
\ee

then we get,

\begin{align}
\begin{split}
E_R(UV) &= \frac{c}{2} \\
E_{R}(IR) &= 0
\end{split}
\end{align}

$E_{R}$ is a UV-finite quantity. Therefore we can see that the renormalized entanglement negativity at least satisfies the asymptotic conditions, i.e, in the UV it is given by the central charge of the theory and in the IR this is zero. The reason that it is going to zero in the IR or in the high temperature limit is that it is an entanglement measure and at very high temperature quantum entanglement goes to zero because the system should crossover to a classical one \cite{Calabrese:2014yza} . This is a non-trivial constraint. Anything that is sensitive to classical correlations may fail to satisfy the IR-condition. Therefore the question is does it satisfy the monotonicity condition, i.e, 

\be
T\frac{d}{dT}\Big|_{L} E_{R} \le 0 \ ?
\ee

If this condition is satisfied then it is a generalized $c$-function. In four dimensions we expect the same thing to happen in the UV. We have to compute the logarithmic negativity for a ball of radius $R$ when the field theory is in a thermal state with temperature $T$. The structure of the UV divergences of the negativity is the same as that of entanglement entropy in the same dimension \cite{Perlmutter:2015vma}. So if we apply the Liu-Mezei operator then we will get a UV finite quantity. The main question is what happens in the IR. Does the renormalized negativity go to zero ? This will be the case if negativity becomes independent of the size of the ball in the high temperature limit. This is a reasonable thing to expect given that there is a finite correlation length of order $\beta$. So we expect the same thing to happen but we cannot prove this right now. It will be fascinating to prove the monotonicity of the negativity at least in two dimensions. 

In the large $c$ limit we expect some simplifications \cite{Faulkner:2013yia}. In fact negativity in the large $c$ limit was considered in \cite{Kulaxizi:2014nma}. Their calculation was for the vacuum sector of the CFT. It will be fascinating to extend the calculation to the thermal state using technology of \cite{Kulaxizi:2014nma, Faulkner:2013yia}. 


Before we end this section we would like to emphasize that we are not saying that the causal horizon entropy density is computing some entanglement measure in a thermal state. That may turn out to be the case but our calculation does not show that. What we can infer from this is the existence of such a monotonic function in field theory which is most likely an entanglement measure. In two dimensional CFT we have shown a potential candidate for this. Causal horizon entropy density represents that quantity in the bulk but perhaps in a different choice of scheme. So numerically they can be different but they will have the same physical content just like in more conventional $c$-theorem.


\subsection{Black Hole Singularity From Loss Of quantum correlation}



There is a different aspect to this problem. Our results can be thought of as a realization of the paradigm that space-time is built out of entanglement \cite{VanRaamsdonk:2010pw}, but in a different setting. In the IR there is no quantum correlation or entanglement because of the effective mass gap in the thermal state. In the bulk our holographic $c$-function is monotonically decreasing and nonzero everywhere except at the curvature singularity. The curvature singularity is the end of space-time and represents the extreme IR of the dual field theory. Therefore the behavior of our $c$-function correlates the two facts : loss of quantum correlation/entanglement in the IR field theory and the end of geometry which in this case is the formation of curvature singularity behind the horizon. In fact this is one of our main motivations for interpreting the $c$-function as an effective bulk measure  of quantum correlation or quantum entanglement between the field theory degrees of freedom at different scales. 

There is another thing which we would like to point out is that since the causal horizon goes behind the black brane horizon and reaches the singularity, the holographic $c$-function is affected by things behind the horizon. Therefore the corresponding boundary $c$-function knows something about physics behind the horizon. If it turns out that the entanglement negativity indeed satisfies the monotonicity condition then this function will have some information about the interior. \footnote{We would like to clarify that we are not talking about a two-sided eternal AdS black hole. We have in mind a black hole, at sufficiently late time, which has formed out of collapsing matter and so the other part of the geometry does not exist. We are making the approximation of a thermal state because the CFT correlators at sufficiently late time are well approximated by thermal correlators.} At infinite temperature when the negativity is zero we are on the singularity because there is no quantum entanglement . As we lower the temperature we are moving away from the singularity but space-time is still very curved because there is only a very small amount of entanglement. So high temperature expansion is an expansion around the singularity. This is a difficult expansion because negativity depends on the full operator content of the theory, bur this may be a virtue of the function for many purposes. 

In \cite{Kraus:2002iv, Fidkowski:2003nf, Festuccia:2005pi} behind the horizon physics was explored using the analytically continued correlation functions in the CFT. The entanglement negativity (or any candidate thermal $c$-function) does not seem to have any simple expression in terms of thermal correlators. It is a highly non-local object. It will be interesting to see if there are more fine-grained characterisations of RG-flow which can tell us about the physics behind the horizon.

\subsection{An Infalling Observer ?}

Let us now go back to the issue of irreversibility associated to a particular geometry. In a black hole geometry there is a natural notion of irreversibility, which is crossing the horizon or falling into the the black hole. Anything that goes into the black hole does not come out. Nothing comes out of the black hole singularity. How is that irreversibility encoded in the field theory ? This is a very difficult question and so we will only try to make a guess. First of all, our $c$-function does not show any particular sharp feature which can be used to predict the existence of horizon. \footnote{We do not know how this picture will change if there is a firewall.} So a natural guess will be that this is a quantity which is associated with an infalling observer. In GR an infalling observer does not see anything special happening while crossing the horizon. So let us make the assumption that the RG-flow or coarse-graining of the thermal state of the CFT describes an infalling observer. We cannot make this statement more precise right now. This assumption together with the fact that this coarse-graining is an irreversible process due to the existence of the $c$-theorem seem to imply that that the observer can never come out of the black hole. The coarse graining starts in the UV when the observer is near the AdS boundary. As we lower the energy scale the observer moves deeper into the bulk. In the extreme IR when the $c$-function hits zero the observer hits the singularity.  This is consistent with the fact that our holographic $c$-function reaches zero at the curvature singularity. Things cannot come out of the black hole singularity because in the field theory there is no unitary RG-flow which starts at $c=0$ and go to $c=a_{UV}$. This is forbidden by $c$-theorem \footnote{We have in mind the generalization of the $c$-theorem to the thermal state. We have proved such a theorem only in the bulk.}. No unitary RG-flow can start at $c=0$ because along the RG-flow $c$ has to decrease. So in RG-time there is an ordering in which the $c=0$ theory always lives in the future. This is also the ordering of time for the infalling observer for whom the black hole singularity is always in the future. This is not quantitative and many things need to be checked before one can say anything conclusive, but at least it is clear that the existence of the $c$-theorem imposes an ordering among different scales in the field theory which, it looks like, can be translated to the bulk under certain assumption and does not immediately produce a contradiction.

\subsection{Tensor Network}

There is another reason to suspect that this may be a correct interpretation. This is related to the tensor network representation of the thermofield double of a scale invariant theory after time evolution. This representation was proposed by Hartman and Maldacena \cite{Hartman:2013qma}. In this picture the tensor network has a scale-invariant UV region and a gapped IR region. The gapped region arises due to the effective mass gap of the thermal state and this represents the interior of the black brane. This resonates well with the behavior of our holographic $c$-function because it shows the extreme thinning of the "effective number of degrees of freedom" near the curvature singularity. A better understanding of this will probably require a more covariant formulation of tensor network ideas. Overall, it seems that MERA \cite{Vidal:2007hda,Swingle:2012wq} might be a proper framework to think about such generalized holographic $c$-functions. The function we have constructed measures the quantum correlation that exists at different scales in the thermal density matrix. MERA does a coarse-graining of the wave function and the generalized $c$-theorem seems to be associated to the irreversibility of that coarse graining procedure. 

\section{Acknowledgements}
We are grateful to Erik Tonni for very helpful correspondence on entanglement negativity.  We would also like to thank Arpan Bhattacharya, Jyotirmoy Bhattacharya and Nilay Kundu for very helpful discussions. The work of SB was supported by World Premier International Research Center Initiative (WPI), MEXT, Japan.

\appendix
\section{Perturbation Near $\lambda\rightarrow 0$}

We have

\begin{eqnarray}
\nonumber z^{2}(\lambda,\eta)\sinh\eta = \frac{1-cn(2\lambda\sqrt{\sinh\eta},1/\sqrt{2})}{1+cn(2\lambda\sqrt{\sinh\eta},1/\sqrt{2})}\\
\Rightarrow z(\lambda,\eta)=\lambda + \frac{\sinh^{2}\eta}{10}\lambda^{5}+\frac{\sinh^{4}\eta}{120}\lambda^{9} 
\end{eqnarray}

where we have kept terms upto order $ \lambda^9 $. Thus we get the solution for $ v(\eta,\lambda) $ as

\begin{equation}
v(\lambda,\eta)=(-1 + \cosh\eta) \lambda - \frac{2}{5} \sinh^4(\eta/2) \lambda^5 - 
 \frac{1}{45} ((-7 + 3 \cosh\eta) \sinh^6\eta/2)\lambda^9
\end{equation}

Using these solutions and the solutions for $ x^i $ we get 

\begin{equation}
c(\lambda,\eta)=1-\frac{\sinh\eta^4}{75} \lambda^8 + .........
\end{equation}

Thus

\begin{align}
\begin{split}
\frac{\partial c}{\partial\lambda}\Big|_{\eta}&= -\frac{8\sinh\eta^4}{75}\lambda^7 \leq 0 \\
c(0,\eta)&=1
\end{split}
\end{align}

\end{document}